\documentclass[pdflatex,sn-nature]{style/sn-jnl}

\usepackage{graphicx}%
\usepackage{multirow}%
\usepackage{amsmath,amssymb,amsfonts}%
\usepackage{amsthm}%
\usepackage{mathrsfs}%
\usepackage[title]{appendix}%
\usepackage{xcolor}%
\usepackage{textcomp}%
\usepackage{manyfoot}%
\usepackage{booktabs}%
\usepackage{algorithm}%
\usepackage{algorithmicx}%
\usepackage{algpseudocode}%
\usepackage{listings}%
\usepackage{siunitx}
\usepackage{lineno}
\usepackage{setspace}
\usepackage{subcaption} %

\begin{document}
\title[Article Title]{Analysis of Collaboration in CS Prizewinning with a Nobel-Turing Comparison}
\author*[1]{\fnm{Boleslaw K.} \sur{Szymanski}}\email{szymab@rpi.edu}

\author[2]{\fnm{Yongtao} \sur{Zhang}}\email{yongtao.zhang@zju.edu.cn}

\author[3]{\fnm{Brian} \sur{Uzzi}}\email{uzzi@northwestern.edu}


\author[1]{\fnm{Mohammed Shahid} \sur{Modi}}\email{modim2@rpi.edu}

\affil*[1]{\orgdiv{Department of Computer Science}, \orgname{Rensselaer Polytechnic Institute}, \orgaddress{\street{110 8th St}, \city{Troy}, \postcode{12180}, \state{NY}, \country{USA}}}

\affil[2]{\orgname{Zhejiang University}, \orgaddress{\street{866 Yuhangtang Road}, \city{Hangzhou}, \postcode{310058}, \country{China}}}

\affil[3]{\orgdiv{Kellogg School of Management}, \orgname{Northwestern University}, \orgaddress{\street{2211 Campus Dr}, \city{Evanston}, \postcode{60208}, \state{IL}, \country{USA}}}

\abstract{In the scientific community, prizes play a pivotal role in shaping research trajectories by conferring credibility and offering financial incentives to researchers. Yet, we know little about the relationship between academic collaborations and prizewinning. By analyzing over 100 scientific prizes and the collaboration behaviors of over 5,000 prizewinners in CS, we find that prizewinners collaborate earlier and more frequently with other prizewinners than researchers who have not yet received similar recognition. Moreover, CS researchers across age groups collaborate more with prizewinners after winning their first prize, and collaborating with prizewinners after their first win increases the likelihood of the collaborator winning an award. We find that recipients of general CS prizes collaborate more than recipients of more specialized prizes, who collaborate less frequently. With Coarsened Exact Matching (CEM) and regression, we find an increase in prizewinning odds with strength of prizewinner collaboration. We examine the context of recent Nobel Prizes going to CS researchers by showing how an increasing share of Physics awards go to Physics-CS collaborations, and contrast Nobel-Turing winning author's trajectories. Our findings shed light on the relationship between prizewinning and collaboration.}




\maketitle

\section{Introduction}\label{sec1}
Scientific prizes are distinctive in the academic communities, shaping research trajectories and influencing academic behaviors. By drawing attention to outstanding research and granting credibility to authors of such works, they transform entire disciplines as well as the trajectories of those scientists~\cite{A2, A25, A5, A3, A1, A4, A27, A34}. Thus, scientific prizes serve as beacons to fields ready for exploration and growth \cite{A6, A7}. Consequently, understanding the role of scientific prizes in scientific communities is crucial. Over the past three decades, there has been a noticeable surge in scientific prizes across numerous disciplines, generating a wealth of data on prizewinners, citations, and collaboration networks worldwide. Together, these datasets enable a comprehensive analysis of the role of prizes in scientific development \cite{B1, B2, B3}.

Scientific prizes positively influence winners' academic careers, contributions, and achievements due to their widespread recognition by institutional authorities and peers. For example, analysis of Nobel Prize winners \cite{A8, A9, A10} is a growing area of research. Such studies have put forth various academic indicators of scholarly potential, such as total citations \cite{A10}, teamwork dynamics \cite{A11, A12, A29}, affiliations \cite{A13}, early and mid-career collaborations \cite{A14, A15, A16} and academic genealogy \cite{A3}. Studies utilizing these indicators have effectively distinguished scholars who have received prestigious scientific prizes in their disciplines from those without such recognition \cite{A17, A30, A19, A20, A31, jin2025}.

These studies often attempt to provide a success paradigm for other scholars' research careers \cite{B4, B5, B6}. While prize-awarding committees frequently choose winners so as to inspire more and better scientific work, it is unclear how winning a prestigious prize affects the future work of the winners themselves. Case studies of Howard Hughes Medical Investigators and John Bates Clark and Fields medalists have shown that prizewinners' papers published before they win their awards gain citations significantly faster than expected of non-winner authors. Moreover, winners increase their chances of winning future prizes \cite{A21}. The presence of the Matthew Effect in these works suggests that accolades and citations are concentrated among a small elite group, especially when they receive early prestigious prizes in their discipline \cite{A22}. These results show that recipients of the Clark Medal and Econometric Society Fellowship increase both productivity (publication count) and impact (citation count) \cite{A23}.

Some findings indicate that Nobel laureates attain fewer citations and publications immediately after winning the prize, but return to pre-award citation rates within four years \cite{B7}. Additionally, after receiving Nobel Prizes, laureates make more significant changes to their research careers compared to their peers \cite{A24}. From a broader perspective, works focusing on the role of scientific prizes in the development of disciplines and the science community find that prizes boost the growth of scientific topics \cite{A4} and push the boundaries of science \cite{A3}.
 
Despite the significant advances made by previous research on the role of scientific prizes in influencing individual awardees and entire disciplines, there remains a gap in our understanding of the relationship between prizewinning and academic collaboration. To address this gap, our study focuses on prizewinners and their collaborators in computer science. 

We collect data on 125 recognized scientific prizes awarded to 5,416 scholars from 1965 to 2022 in computer science from various sources. We link these prizes with authors in the large 'SciSciNet' dataset \cite{C1} enabling us to build a network of prizewinner's collaboration networks. This dataset serves as the base for our work. For a full data description, refer to SI Section 1. We build enhanced networks with more data and parameters for various analyses, the details of which are provided in their corresponding sections.

We find that prizewinners in computer science collaborate more frequently and earlier with other prizewinners than researchers who have not yet received similar recognition. Compared to their pre-prize academic careers, scholars in computer science tend to collaborate more with other prizewinners after receiving a prize. We observed this trend across different age groups of prizewinners. 

To investigate the impact of prizewinning on collaborators, we focus on their future academic development at various career stages. Specifically, we divide the collaborators of these prizewinners into pre- and post-prizewinning collaborations. By examining the award outcomes of these two groups over time, we find that a more significant portion of post-prizewinning collaborators will become prizewinners in the future. We also investigate the collaboration patterns among specific prizes, revealing that recipients of general prizes in computer science are more likely to collaborate than those of specialized prizes. Additionally, we observed that specialized prizes' prizewinning cycles are shorter than general prizes.

Next, we investigate the strength of one's collaboration with a prizewinner as a factor in eventually winning a prize. We use various parameters from the SciSciNet database to describe prizewinners and their collaborators. We use Coarsened Exact Matching (CEM) to create balanced author groups for those parameters and fit them to a regression model. Our models show that authors with strong collaborations with existing prizewinners are more likely to win awards themselves, both throughout their whole career and before they win their first prize, when compared to authors with weaker collaborations.

Finally, we analyze Turing Award and Nobel Prize winners in particular. For one, the 2024 Nobel Prize being awarded to CS researchers could be part of a broader trend of computer scientists winning a larger share of academic prizes even outside of CS. Thus, we examine many field-specific awards in science, contrasting them to Computer Science-specific awards. We use a modern Large Language Model API to classify award instances as single-field or inter-disciplinary/multi-disciplinary based on the authors and works that they were awarded to. We find that inter and multi-disciplinary CS authors, and works with a CS component, have gained a greater share of out-discipline awards over time. Secondly, we contrast the careers of scholars who won Turing Awards to those who won Nobel Prizes in Physics, Chemistry and Medicine to show how the most accomplished CS researchers differ from those in other fields.

Our findings contribute to a deeper understanding of the interplay between prizewinning and academic cooperation and offer practical guidance for enhancing the effectiveness of scientific collaborations, which, in turn, can drive future advancements in computer science and potentially other fields, ultimately contributing to the broader progress of scientific and technological endeavors.
 
\section{Results}
\subsection{Academic collaboration as an early signal of prizewinning}
\begin{figure}[!t] 
\centering
\includegraphics[width=4in]{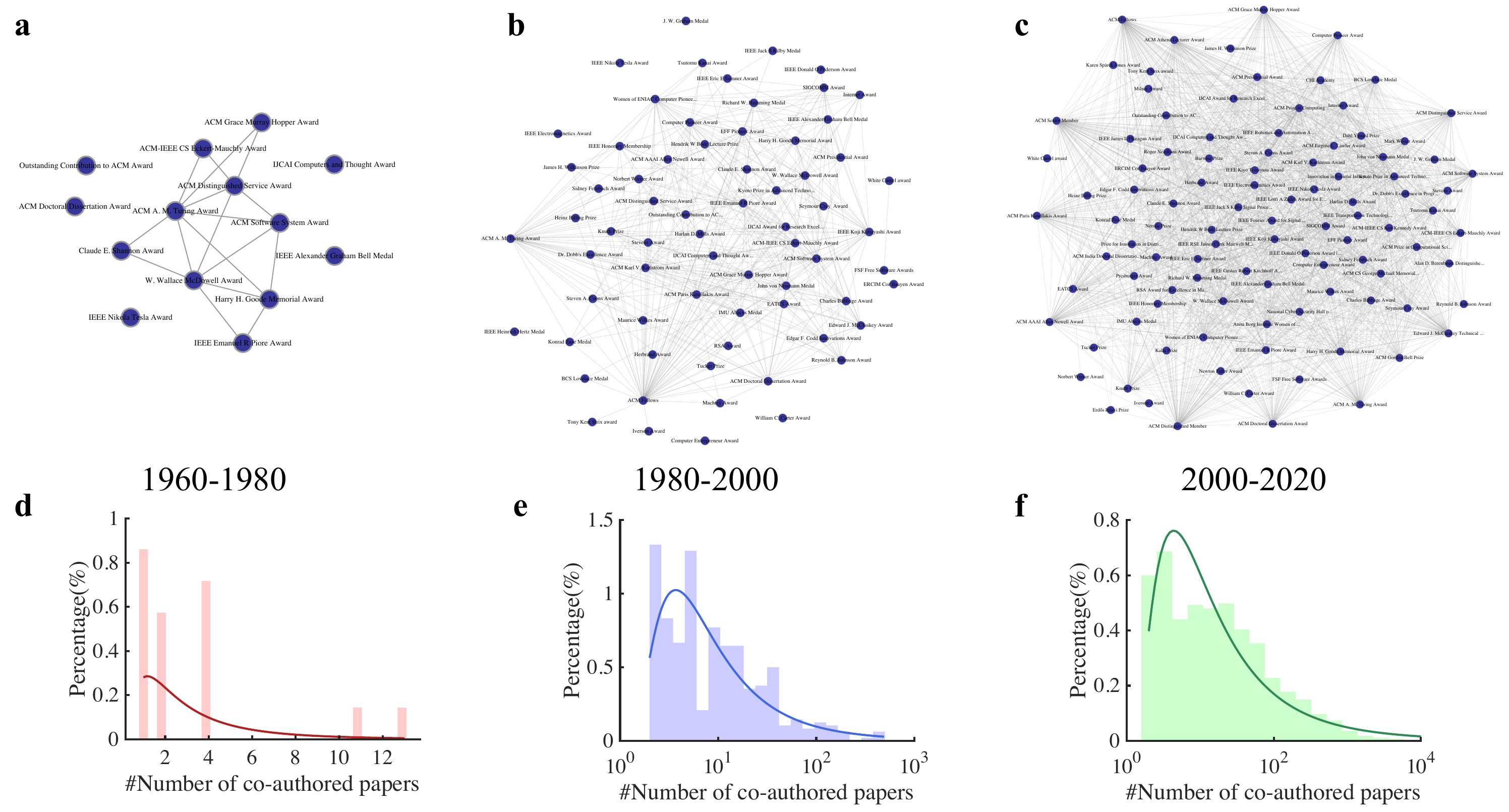}
\caption{\textbf{Evolution of scientific prize network in computer science. A}, Visualizations of the collaboration networks among scientific prizes in computer science across three distinct periods: 1960-1980 in \textbf{a}, 1980-2000 in \textbf{b}, and 2000-2020 in \textbf{c}. Each node represents the awardees of a specific scientific prize during the corresponding period. Edges between nodes indicate the co-author relationships between groups of awardees from different prizes. \textbf{d-e}, Probability distribution functions illustrate the number of co-authored papers among all possible prizewinner pairs within each period.}
\label{F:1}
\end{figure}

We focus on prizewinners across different periods and investigate whether their collaboration behaviors have changed significantly over time. We visualize the evolving landscape of collaboration among these elite researchers in Figure \ref{F:1}a-c. Each node in the network represents a different prize, while edges indicate co-author relationships between winners of various prizes within the same period. 

The number of prizes has increased over time, and the collaboration network has become much denser in the last two decades, indicating that in recent years prizewinners have a stronger preference for collaborating with other prizewinners. While nearly one-third of the prizes in the early years had winners with no academic collaborations with other prizewinners, all recent prizes have some winners who have collaborated with other winners to some extent. We examine the number of co-authored papers between prizewinners across different periods to provide a more quantitative perspective on this trend. 

\begin{figure}[!t]
\centering
\includegraphics[width=3in]{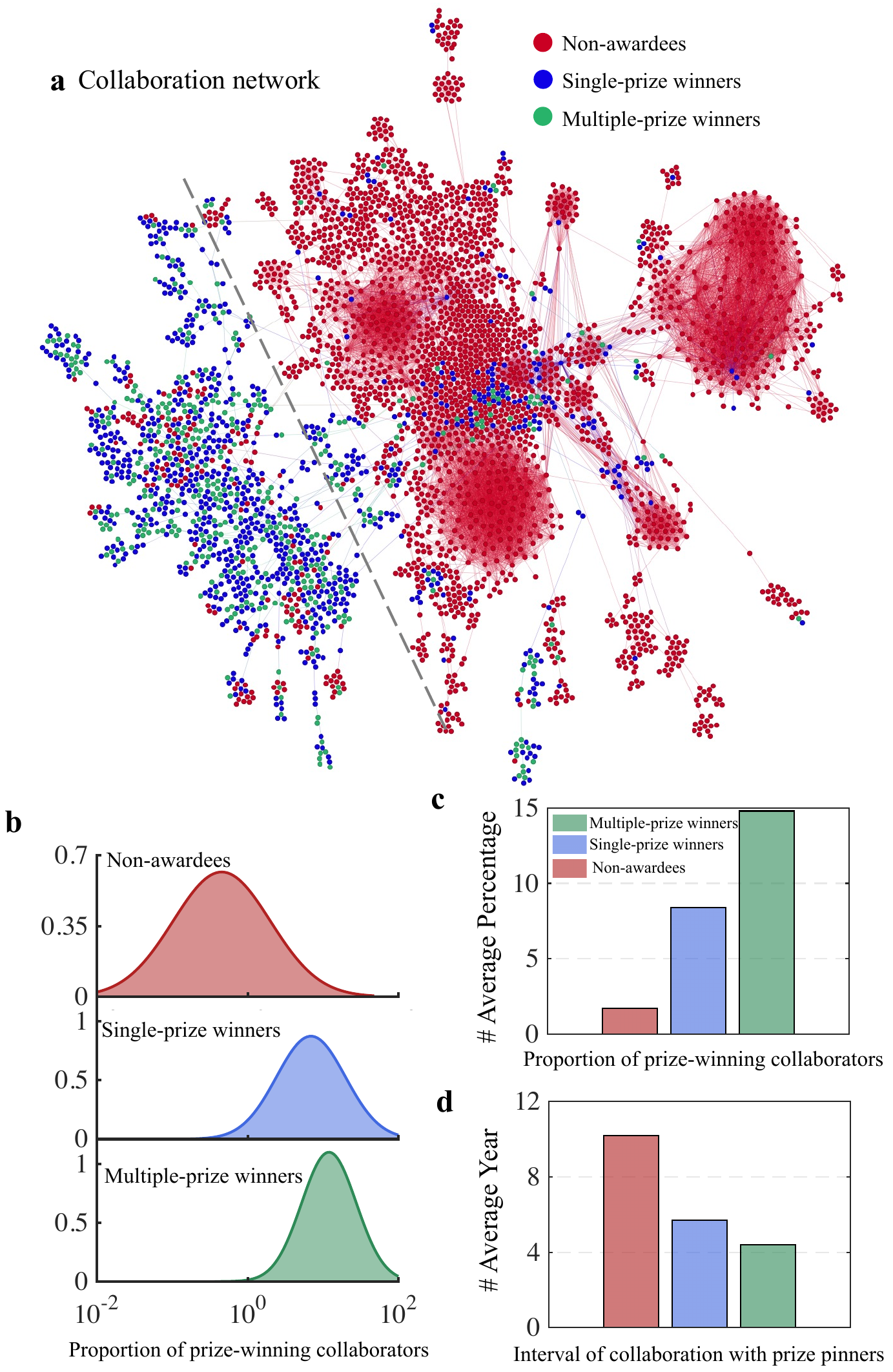}
\caption{\textbf{Characterizing the collaboration patterns of computer science scholars.} \textbf{a.} The collaboration network among three types of computer science scholars. We remove edges with weights less than ten and the resulting isolated nodes to make the network readable. \textbf{b.} The density distribution of prizewinning collaborators as the fraction of the number of prizewinning computer science collaborators to the total number of collaborators. For winners, we only consider collaborators they had before receiving their first prize. \textbf{c.}  The average percentage values of the fraction of prizewinning collaborators. \textbf{d.} The average time intervals between first publication year and first year of collaboration with a prizewinner.}
\label{F:2}
\end{figure}

As shown in Figure \ref{F:1}d-f, the probability distribution functions demonstrate a proliferation in the number of co-authored papers, with significant differences observed across the periods. The observed differences support the observation of increased collaboration density within the prizewinning community and suggest that academic collaboration, particularly with other prizewinners, may serve as an early indicator of future awards and recognition.

\subsubsection{Collaboration Patterns}
We characterize collaboration behaviors and explore distinct characteristics among prizewinners. To do this, we categorize researchers in computer science into three groups based on their prizewinning status: winners of multiple, single, and no prizes. To ensure a robust comparison, we define non-awardees as researchers who have been inactive for at least ten years (i.e., no publications since 2010) and have a substantial publication record of at least 500 citations. We randomly sample 5,000 scientists from this subset to serve as baseline comparisons. 

As shown in Figure \ref{F:2}a, we visualize the collaboration network among these three groups, with nodes color-coded to represent their winning status. An edge between two nodes indicates that the two scientists have at least co-authored one paper. The figure reveals two disjoint clusters, one comprising non-awardees and another including prizewinners. For winners of multiple and single prizes, we specifically consider the collaborators they had before receiving their first prize. Figure \ref{F:2}b illustrates the probability distribution function of the fraction values for each group. Consistent with the observations in Figure \ref{F:2}a, prizewinners exhibit a significantly higher fraction value than non-awardees, indicating that the prizewinners have a more significant fraction of other prizewinners among their collaborators, particularly for winners of multiple prizes. 

We present each distribution's average percentage value in Figure \ref{F:2}c. From the figure, numerous prize winners exhibit the highest average fraction of about 15\%. Single-prizewinners follow closely, with a fraction of approximately 9\%. In contrast, non-awardees have the lowest fraction, with a value of less than 2\%. These findings suggest that computer scientists collaborating more frequently with prizewinners increase their probability of winning. 

To investigate collaboration as an early signal of prizewinning, we analyze the collaboration patterns among these three groups over time, examining how they influence future prizewinning at different stages of an academic career. Figure \ref{F:2}d depicts the time interval between each scholar's first publication and the first collaboration with a prizewinner. Interestingly, this trend diverges from the overall fraction of prizewinning collaborators observed in Figure \ref{F:2}c. Winners of multiple prizes exhibit the shortest interval between their first publication and first collaboration with a prizewinner, followed by a single prize winner. Non-awardees, on the other hand, require the longest time interval. This observation suggests that the timing of a scholar's first collaboration with a prizewinner can also serve as an early indicator of future potential for winning prizes. Specifically, computer scientists collaborating with prizewinners earlier in their careers are more likely to receive greater recognition and awards.
 
\subsection{Prize transition in computer science}
How can academic collaboration be optimized to increase the likelihood of winning prizes? To investigate this, we examine the link between computer science prize trends and author collaboration/winning patterns. We formulate a probability adjacency matrix of the prize network to characterize this relation. In this matrix, $\textbf{A}={A_{ij}}$, the element $(i, j)$ is defined as:

\begin{equation}
A_{ij}=\frac{P(\text{winner of prize } j \mid \text{collaborator of prize } i)}{P(\text{winner of prize } j)}
\label{Eq:1}
\end{equation}

where $P(\text{winner of prize } i)$ is the probability of randomly selecting a scholar who is a winner of prize $i$. The scholars considered here include all prizewinners and their collaborators. The conditional probability $P(\text{winner of prize } j \mid \text{collaborator of prize } i)$ represents the likelihood that collaborators of prize $i$ winners win prize $j$ in the future. Therefore, $A_{ij}$ indicates the strength of the association between winning prize $j$ and collaborating with the prize winners $i$. If the events of collaborating with winners of prize $i$ and winning prize $j$ are statistically independent, $A_{ij} = 1$. A value of $A_{ij}$ greater than 1 suggests that collaborations with winners of prize $i$ positively influence winning prize $j$. Specifically, the higher the $A_{ij}$ is, the stronger the evidence that such collaborations significantly increase the likelihood of winning prize $j$.

\begin{figure}[t!]
\centering
\includegraphics[width=4in]{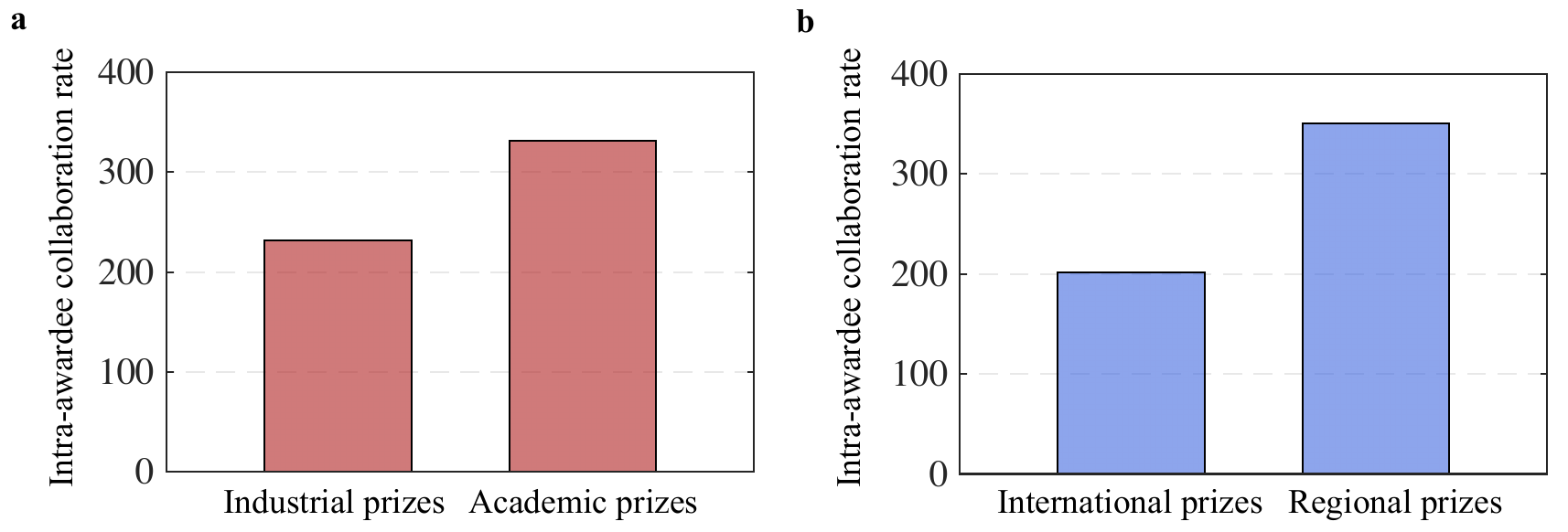}
\caption{\textbf{Intra-awardee collaboration rate. a},  Comparison of intra-awardee collaboration rates for industrial and academic prizes. The lower rate for industrial prizes suggests that collaborators of academic prizewinners have a higher probability of subsequently winning academic prizes. \textbf{b}, Comparison of intra-awardee collaboration rates for international and regional prizes. The higher rate for regional prizes indicates that internal collaboration among winners of regional prizes is correlated with higher chances of winning the same prize in the future.}
\label{F:3}
\end{figure}
 
We apply Eq. (\ref{Eq:1}) to 105 computer science prizes categorized into two groups: industrial and academic. Industrial prizes, often sponsored by commercial entities, emphasize innovation with direct industrial application and commercial viability (e.g., the IEEE Internet Award) \cite{A32}. Academic prizes sponsored by educational and research institutions recognize contributions to the broader development of computer science (e.g., the Turing Award). 

We compute the pairwise conditional probability $A_{ij}$ for each prize pair, which shows the impact of intra-awardee collaboration, where collaborators subsequently win the same award. For the diagonal elements ($A_{ii}$) of the conditional probability matrix, higher values indicate a more substantial positive correlation of collaborating with a past winner of prize $i$ on winning prize $i$ in the future. Figure \ref{F:3}a reveals a lower intra-awardee collaboration rate for industrial prizes than academic prizes. This difference in the collaboration rate suggests that collaborators of academic prizewinners are more likely to win future prizes. We also categorize prizes by scope, including international and regional. International prizes recognize achievements globally, while regional prizes focus on specific geographic areas (e.g., the Hong Kong ICT Awards). Figure \ref{F:3}b shows a higher intra-awardee collaboration rate for regional prizes than international prizes. The observed collaboration rates imply that winners of regional prizes are more likely to engage in internal collaboration, and this collaboration correlates with a higher likelihood of them winning the same prize again. This may be due to repeat collaboration dynamics that have also been observed among Nobel Prize awardees \cite{A33}. Refer to SI Section S2 for a visualization of the adjacency matrix as a heatmap with additional results related to career age of prizewinners and notability of prizes (SI Figure S1).
 
\begin{figure}[t!]
\centering
\includegraphics[width=5in]{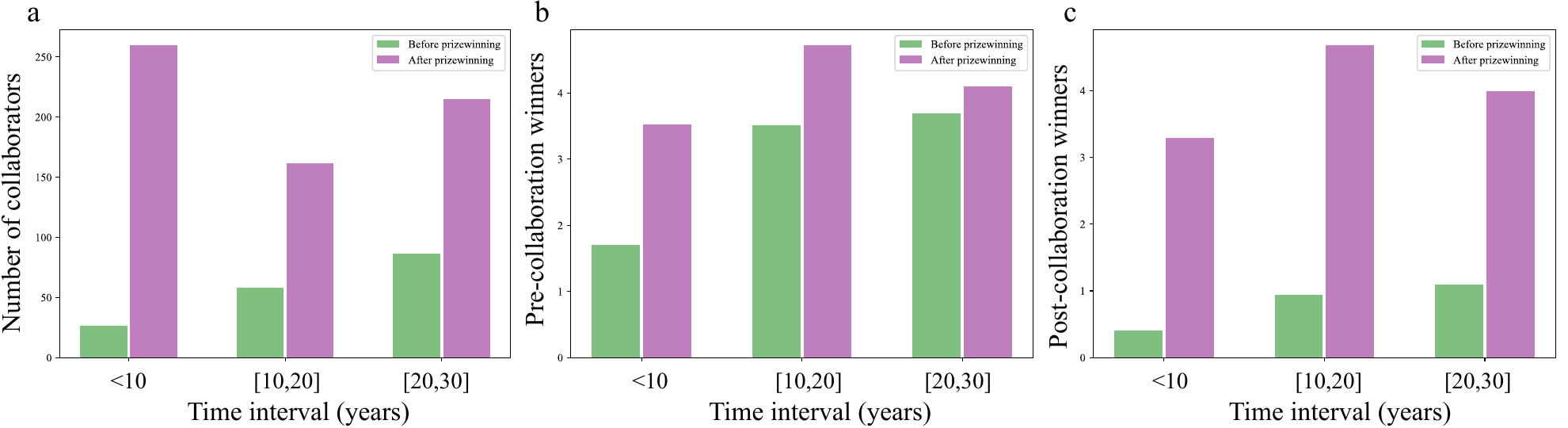}
\caption{
\textbf{Effect of prizewinning on collaboration pattern. A}, Average number of collaborators of winners of prizes before and after receiving their award. The X-axis represents the career stages of prizewinners,  categorized into three groups based on the time from their first publication to winning the first prize: under 10 years, 10 to 20 years, and over 20 years. The Y-axis quantifies the average number of collaborators. Green bars depict the average value before prizewinning, and purple bars show the average after prizewinning. \textbf{b}, Comparison of the number of prizewinning collaborators. Green and purple bars correspond to the collaborator's pre- and post-prizewinning. \textbf{c}, Future award outcomes of collaborators within ten years following their collaboration with a prizewinner.}
\label{F:5}
\end{figure}
 
\subsection{Effect of prizewinning on collaboration pattern}
We investigate if winning a prize alter a scholar's preference for collaborating with fellow awardees versus non-awardees and whether collaboration with a prizewinner increases the likelihood of prizewinning. To investigate these questions, we analyze the careers of prizewinners by dividing them into two phases: pre-award and post-award. We assess the impact of winning a prize on the award status of collaborators during these phases. As illustrated in Figure \ref{F:5}, we depict the career stages of prizewinners with green and purple bars representing the pre- and post-prizewinning periods, respectively. 

As shown in Figure \ref{F:5}a, there is a significant increase in the average number of collaborators post-award across all categories. This increase is nearly tenfold for early-stage researchers, rising from an average of 25 collaborators pre-award to over 250 post-award. In contrast, the increase is from approximately 80 to 210 collaborators for those in later career stages. This pattern indicates that while the stage of a scholar's career at the time of the award influences the magnitude of change in collaboration numbers, the overall trend toward increased collaboration after winning prizes is consistent. 
 
Figure \ref{F:5}b details the shifts in collaboration preferences among prizewinners, explicitly focusing on their propensity to collaborate with other awardees before and after winning their prize. The green histograms represent the average number of prizewinners among the collaborators before the award, while the purple histograms show this metric post-award. Notably, early-career prizewinners exhibit a more significant increase in collaborations with other awardees post-award than those in later careers. Although late-career prizewinners tend to collaborate more with other awardees after their recognition, the difference is relatively modest, indicating that the increase in collaboration with fellow awardees is consistent regardless of the career stage. This figure demonstrates that prizewinners, irrespective of their academic stage, are more inclined to collaborate with other awardees after receiving their awards. 

Furthermore, we assess the future award outcomes of collaborators within a decade following their collaboration with the prizewinner. Our findings reveal that, while a portion of collaborators receive awards post-collaboration, those who collaborated post-prizewinning exhibit a significantly higher likelihood of winning awards within the next ten years than those who collaborated before winning any prizes. Figure \ref{F:5}a-c collectively suggests that winning a prize is associated with a shift in the scholar's collaborative tendencies, increasing their interactions with other awardees, and significantly enhancing their collaborators' academic careers. Such a shift is more likely to increase collaborations undertaken after a scholar's recognition, which positively correlates with their collaborators' subsequent award prospects.

\subsubsection{Analysis of Collaboration Strength and Prizewinning Probability using Coarsened Exact Matching}
To better understand the effect of prizewinner collaboration strength on the probability of prizewinning, we use a two-prong approach of the Coarsened Exact Matching (CEM) to balance author groups and logistic regression to test hypotheses on those groups. We consider well-balanced matched groups to be those with an average Standardized Mean Difference $SMD<0.1$ by convention. Once the groups are created, we fit a weighted logistic regression model with the covariates included; see the ``Methods'' section for more details on the covariates chosen and techniques used.

\begin{figure}[t!]
    \centering
    \begin{subfigure}[t]{0.4\textwidth}
        \centering
        \includegraphics[width=\linewidth]{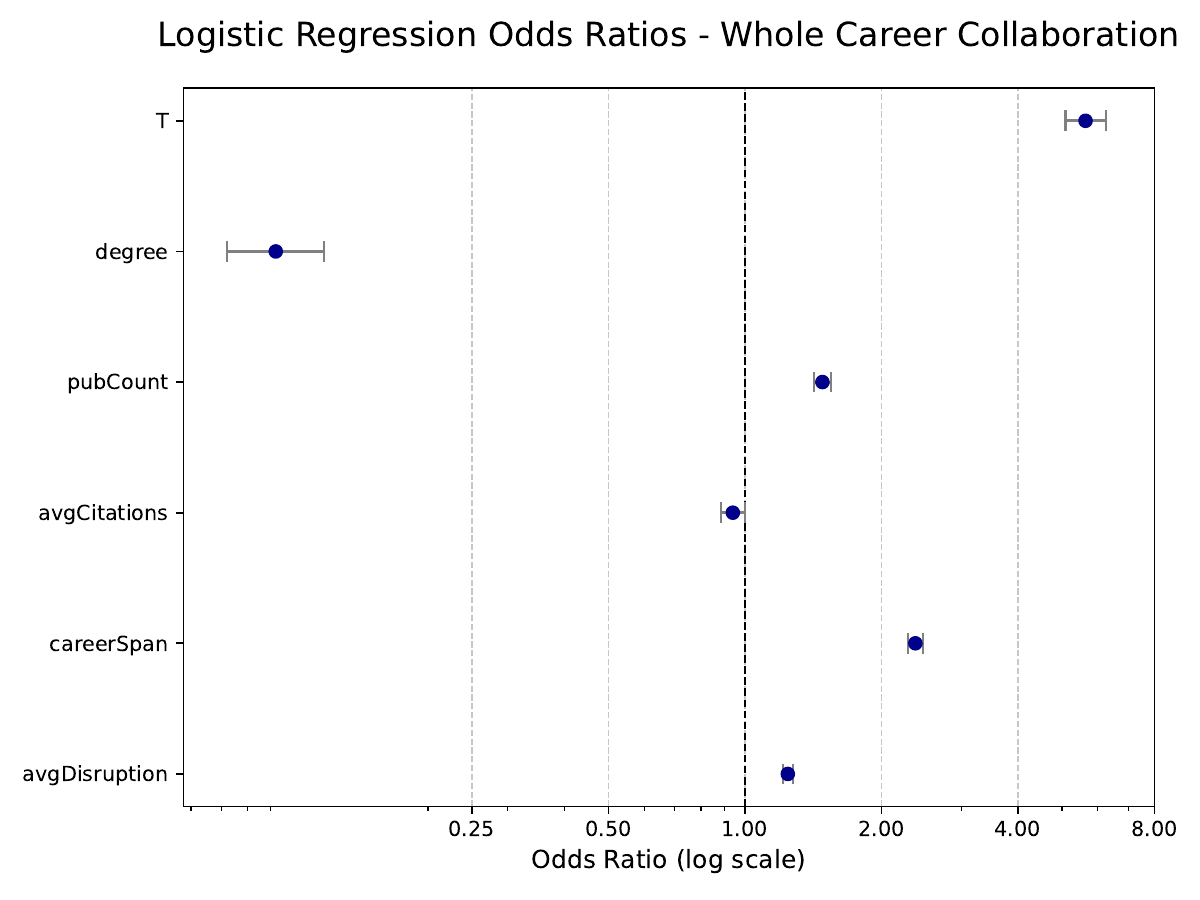}
        \caption{}
        \label{fig:forest_whole}
    \end{subfigure}%
    \hspace{1cm}
    \begin{subfigure}[t]{0.4\textwidth}
        \centering
        \includegraphics[width=\linewidth]{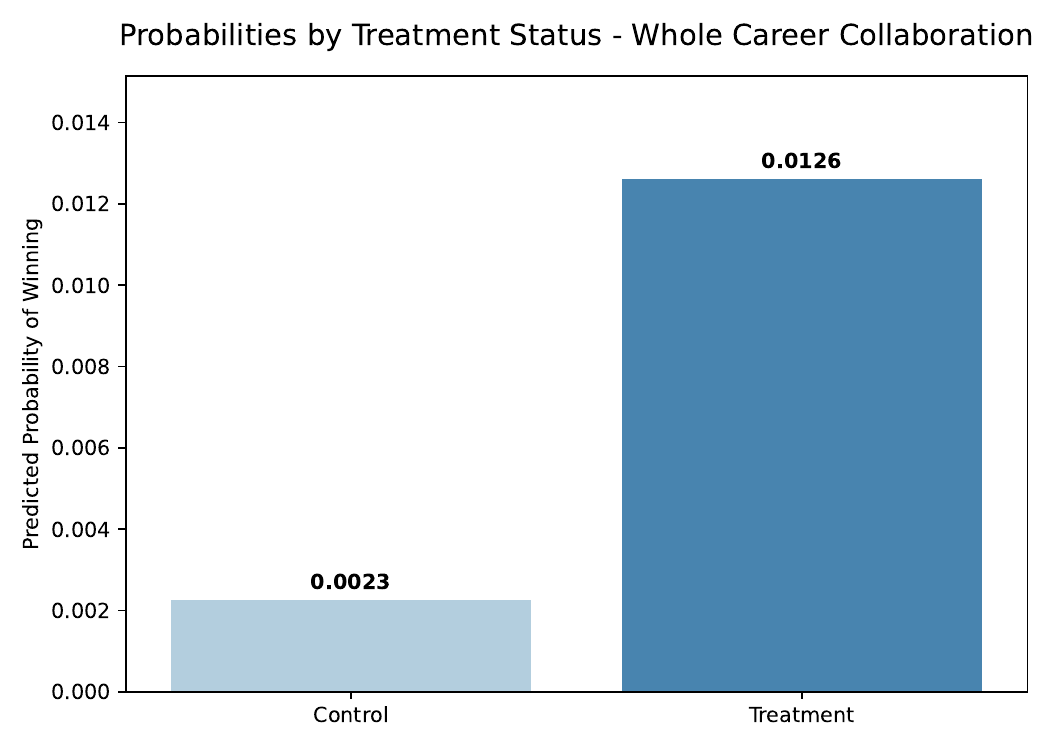}
        \caption{}
        \label{fig:prob_whole}
    \end{subfigure}
    
    \caption{Results of logistic regression on the whole-career collaboration objective with treatment (T) being the number of co-authored papers with a prizewinner. \textbf{(A)} Forest plot of odds ratios for each covariate and treatment. \textbf{(B)} Predicted probabilities of each covariate.}
    \label{fig:whole_plots}
\end{figure}

We pursue two objectives with this analysis, \textbf{(i)} to estimate the effect of strong collaboration between a prizewinner-author pair over their whole careers on the beneficiary author's odds of being a prizewinner; \textbf{(ii)} to estimate the effect of strong collaboration between a prizewinner-author pair before the beneficiary author won their first prize on the odds of winning that prize. 

After matching, we obtain a weighted Average Treatment Effect (ATE) of $0.0198$ for the whole-career collaboration sample corresponding to a $1.98$ percentage point increase in the odds of winning an award in an author's career when that author has a high number of shared papers with a different prizewinner compared to control. Similarly, for the pre-award collaboration objective, we get a weighted ATE of $0.017492$ ($1.7492$ percentage point increase). Note that while the ATEs appear to be small, prizewinning is a rare event and we discuss the relatively high increase they represent in the subsections below. (Refer to SI Section S3 for all technical details, including Table S2 and Table S3.)

\subsubsection{Analysis of Strong Whole-Career Collaboration}

 \begin{figure}[t!]
    \centering
    \begin{subfigure}[t]{0.4\textwidth}
        \centering
        \includegraphics[width=\linewidth]{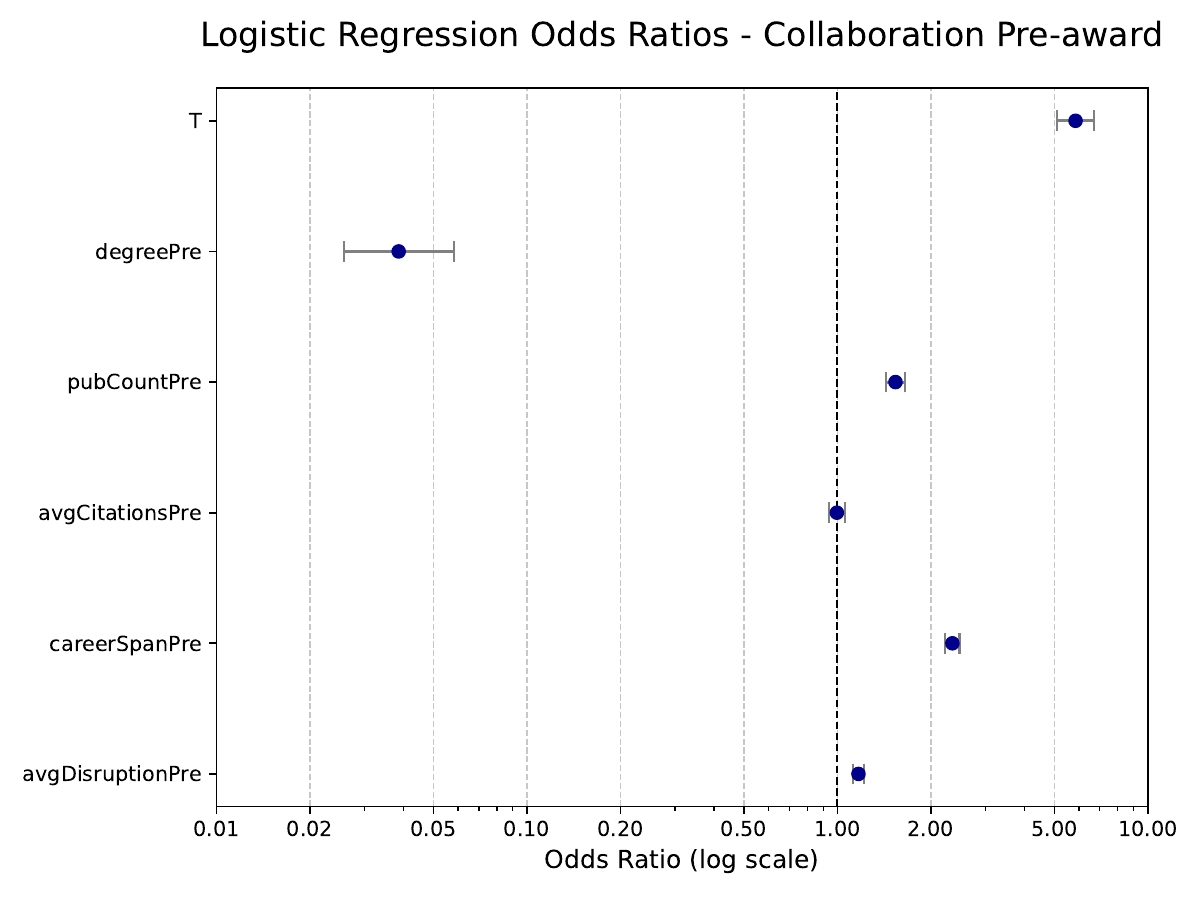}
        \caption{}
        \label{fig:forest_preaward}
    \end{subfigure}%
    \hspace{1cm}
    \begin{subfigure}[t]{0.4\textwidth}
        \centering
        \includegraphics[width=\linewidth]{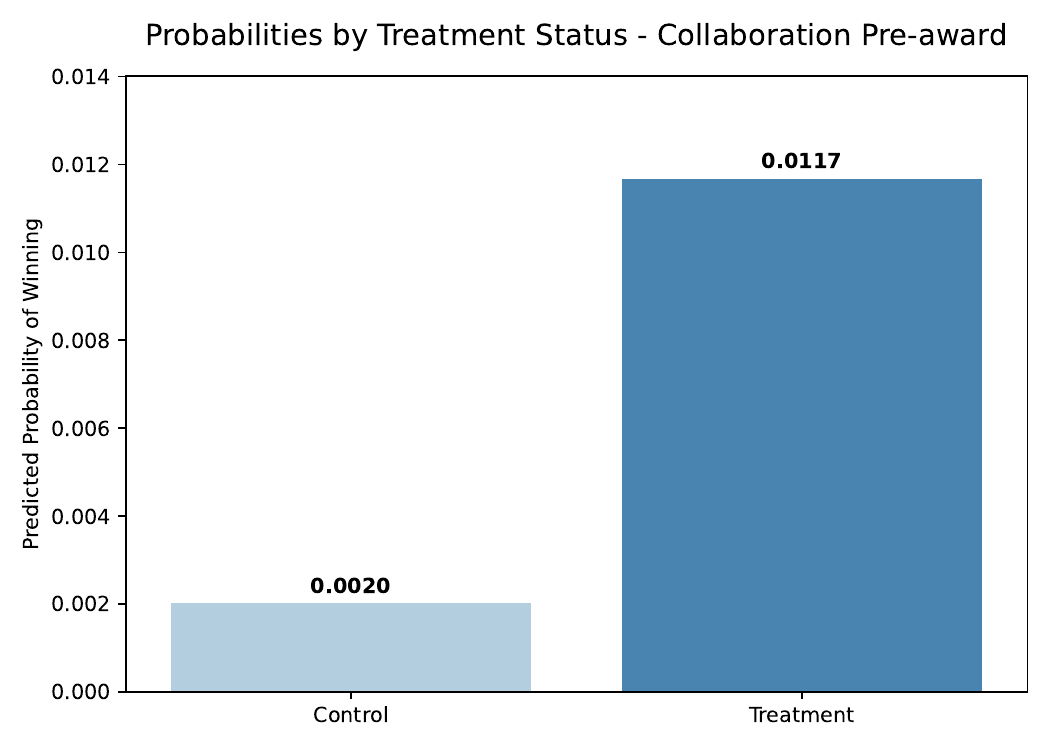}
        \caption{}
        \label{fig:prob_preaward}
    \end{subfigure}
    \caption{Results of logistic regression on the pre-award collaboration objective with treatment (T) being the number of co-authored papers with a prizewinner before winning a prize. \textbf{(A)} Forest plot of odds ratios for each covariate and treatment. \textbf{(B)} Predicted probabilities of each covariate.}
    \label{fig:preaward_plots}
\end{figure}

We fit a weighted logistic regression model to the matched samples for the whole-career collaboration objective. The regression results reveal that the treatment is a strong positive indicator of prizewinning as visualized by the forest plot of odds ratios shown in Figure \ref{fig:whole_plots}(a). Apart from the treatment, publication count, career span and average disruption are positively associated with prizewinning in descending order of magnitude, while degree is negatively associated and average citations are as well. Note that average citations was found to be statistically insignificant with $p=0.049$ with all other covariates having $p<0.001$. Figure \ref{fig:whole_plots}(b) captures a jump of over five-fold in the predicted probability of prizewinning between the treatment and control (Odds Ratio $5.638$). The negative association of degree (number of co-authors) to prizewinning could be because working with a larger number of authors would prevent a scholar from building a strong relationship with a single author. Having a longer career and a large number of publications appears to be more important to prizewinning than producing more 'disruptive' works. The Average citations factor is not statistically significant which may be because the model found it to be too similar to another factor, thus diluting its impact.

\subsubsection{Analysis of Strong Pre-award Collaboration}

We repeat the weighted logistic regression following the same process as before, this time for the pre-award collaboration objective, with the regression outcomes for each attribute captured in Figure \ref{fig:preaward_plots}(a). This outcome reveals a more significant effect between having a strong collaboration with a prizewinner and winning a prize. The treatment plays a bigger role in the outcome than before, causing a nearly six-fold increase (Odds ratio $5.849$) in the odds of prizewinning. This is a stronger result in terms of predicted probability relative to the first experiment, with the six-fold increase visualized in Figure \ref{fig:preaward_plots}(b). Degree is negatively associated with prizewinning here as well, suggesting that prior to prizewinning, having a large co-author network hobbles scholars' relationships with prizewinning mentors.

\begin{figure}[t!]
\centering
\includegraphics[width=3in]{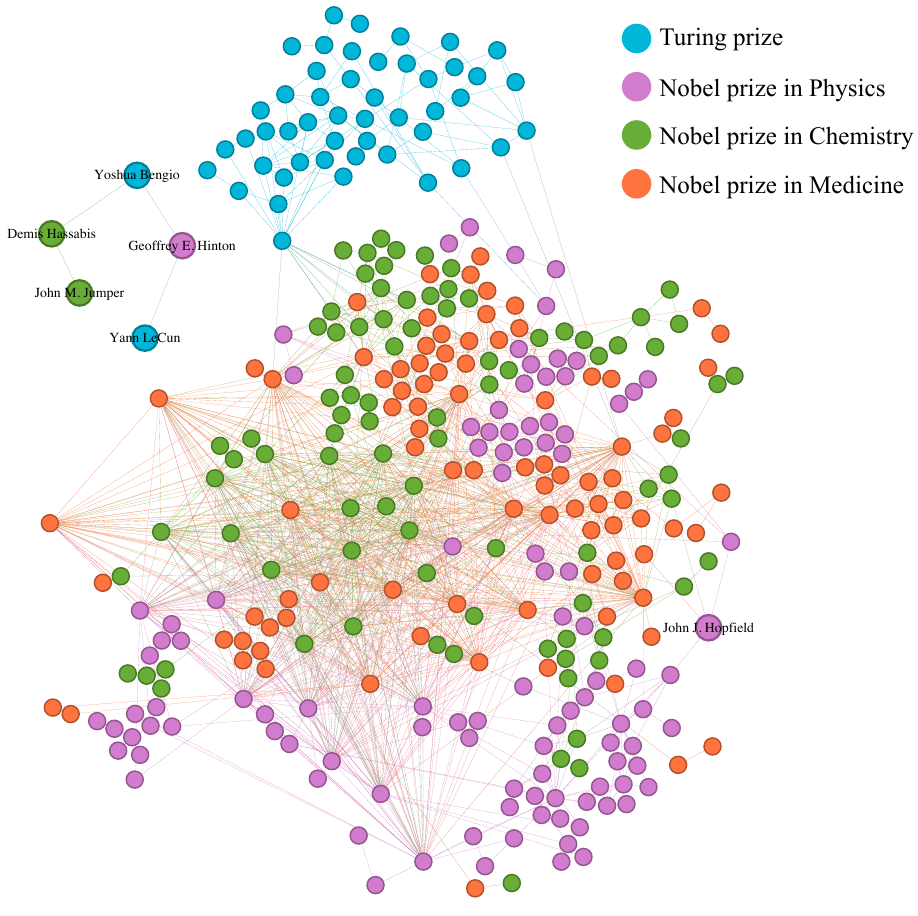}
\caption{\textbf{Collaboration network among Turing and Nobel Prize winners.} This network visualizes the academic collaborations among Turing Award and Nobel Prize winners. Each node represents a prize winner and is color-coded according to the prize(s) they have received. Edges indicate co-authorship relationships between two prize winners. For clarity, isolated nodes representing prize winners without co-authorship with others are not displayed.}
\label{F:6}
\end{figure}

\subsection{Unprecedented Nobel recognition of computer scientists}
The 2024 Nobel Prize committee awarded the Nobel Prizes in Physics and Chemistry to four computer scientists, causing a stir in the academic community. Specifically, Geoffrey E. Hinton and John J. Hopfield received the Nobel Prize in Physics for their contributions to the artificial neural networks in machine learning, while Demis Hassabis and John M. Jumper were awarded the Nobel Prize in Chemistry for their work on computational protein design. This extraordinary historical event prompts us to take a closer look at the interdisciplinary integration of different fields in recent years, especially the collaboration between computer science and other disciplines in the context of prizewinning.

To understand the uniqueness of these four computer scientists regarding academic collaborations, we construct a collaboration network among Turing and Nobel Prize winners, as shown in Figure \ref{F:6}. The nodes in the network are color-coded to represent different prizes, with the four most recent Nobel Prize winners highlighted with larger nodes and labeled with their names. From the figure, we observe limited connections between Turing Award and Nobel Prize winners; specifically, most Turing Award winners do not have close connections with Nobel laureates, with a few exceptions. The three most recent Nobel Prize winners focused on AI -- Demis, John, and Geoffrey -- have collaborated with each other, along with two other Turing Award winners, Yoshua Bengio and Yann LeCun. These five form a sub-collaboration network that is isolated from other prize winners. This indicates that research in AI and the widespread use of AI technology in different fields have provided these top scientists with the opportunity to win the highest honors in other domains. This prompts us to further ask whether other emerging topics in computer science, and their researchers, might also have a chance to win a Nobel Prize in the future.

\subsubsection{Share of non-CS research prizes granted to Computer Science}
\begin{figure}[t!]
    \centering
    \begin{subfigure}[b]{0.32\textwidth}
        \includegraphics[width=\textwidth]{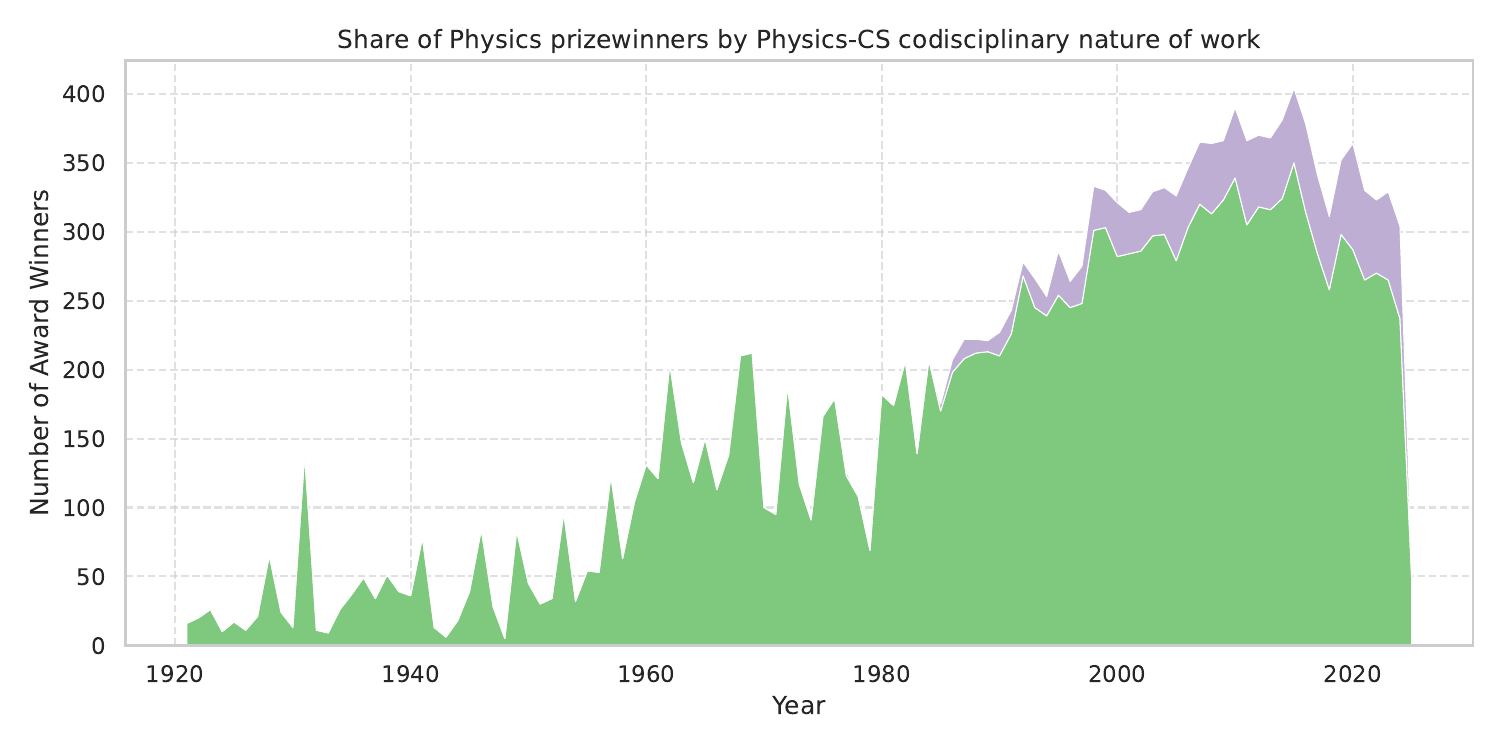}
        \caption{Share with CS}
        \label{fig:share_phy_cs}
    \end{subfigure}
    \hfill
    \begin{subfigure}[b]{0.32\textwidth}
        \includegraphics[width=\textwidth]{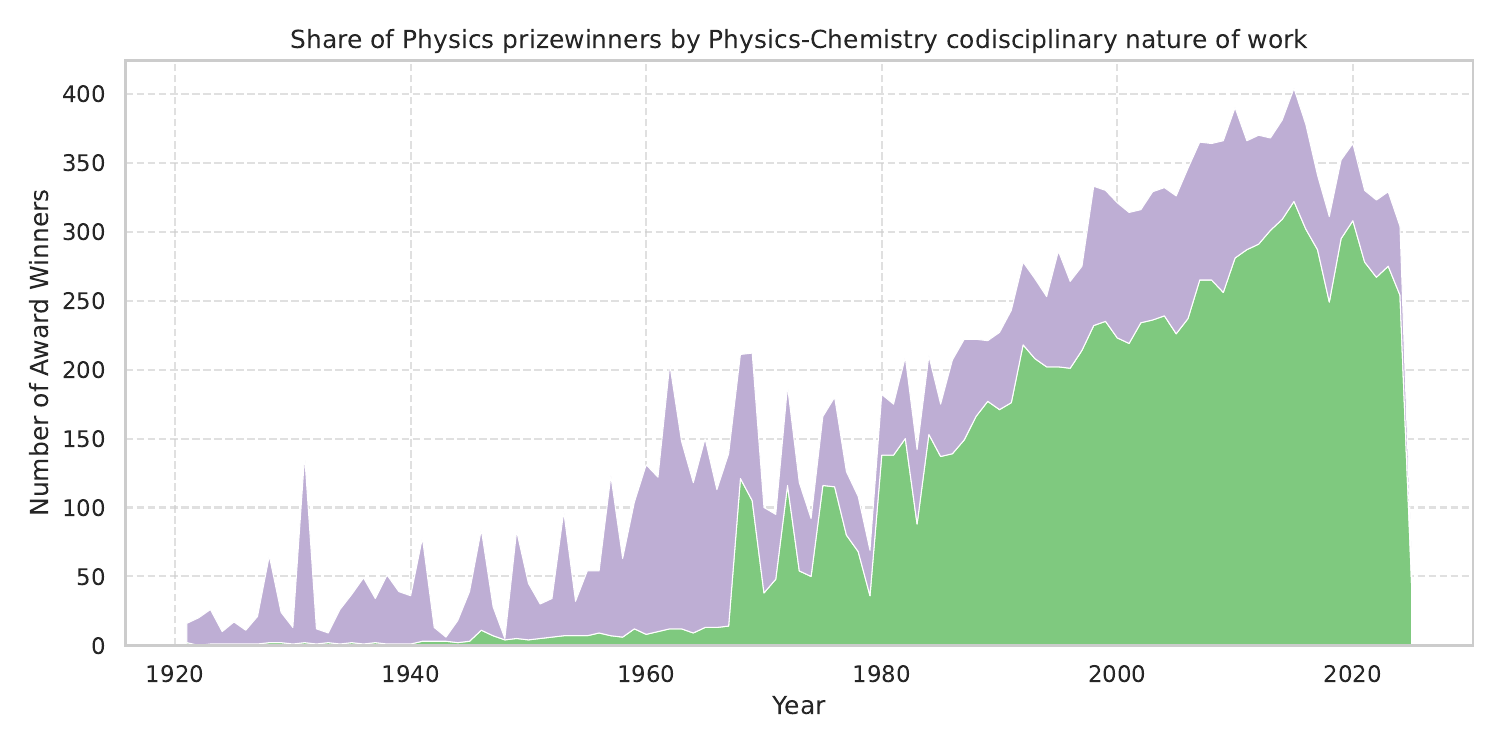}
        \caption{Share with Chemistry}
        \label{fig:share_phy_chem}
    \end{subfigure}
    \hfill
    \begin{subfigure}[b]{0.32\textwidth}
        \includegraphics[width=\textwidth]{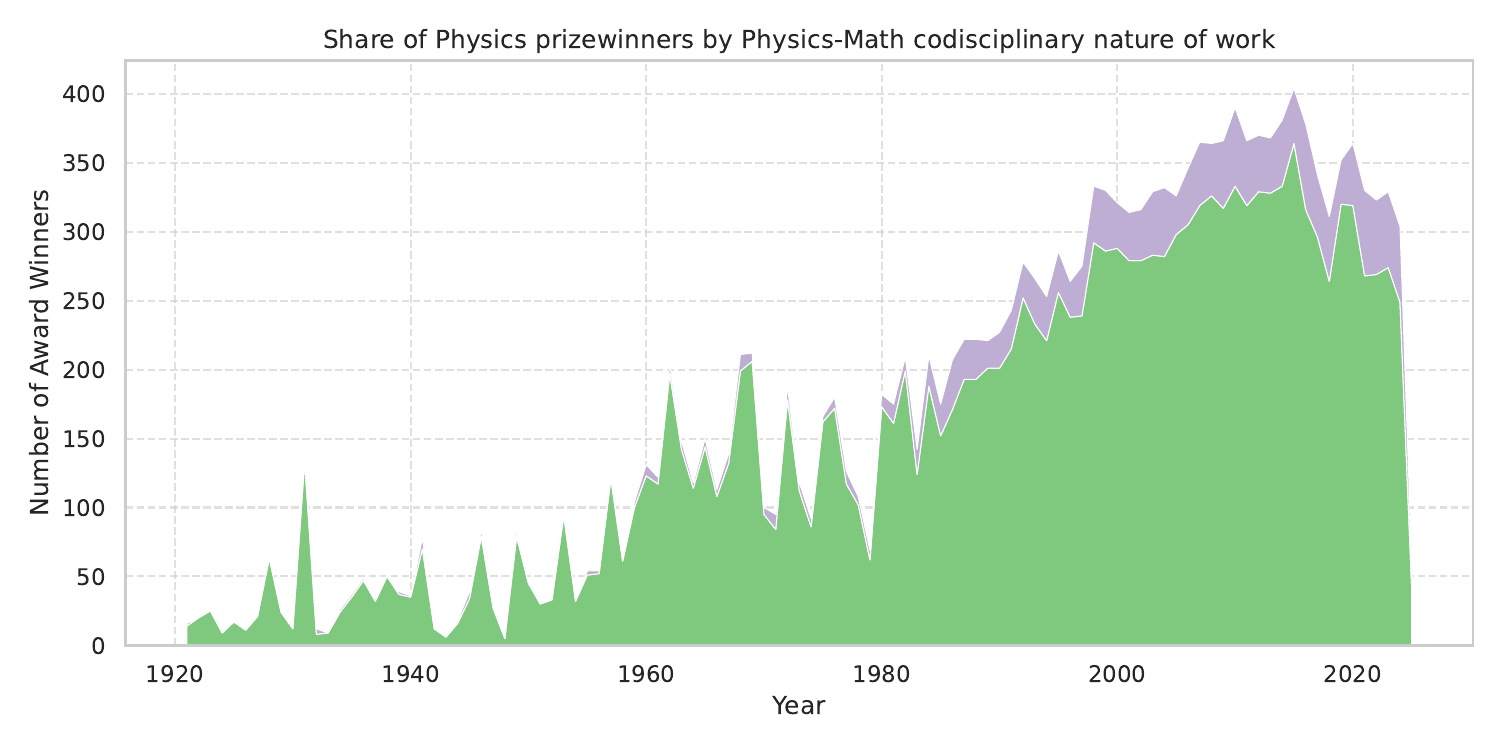}
        \caption{Share with Math}
        \label{fig:share_phy_math}
    \end{subfigure}

    \vspace{1em}

    \begin{subfigure}[b]{0.32\textwidth}
        \includegraphics[width=\textwidth]{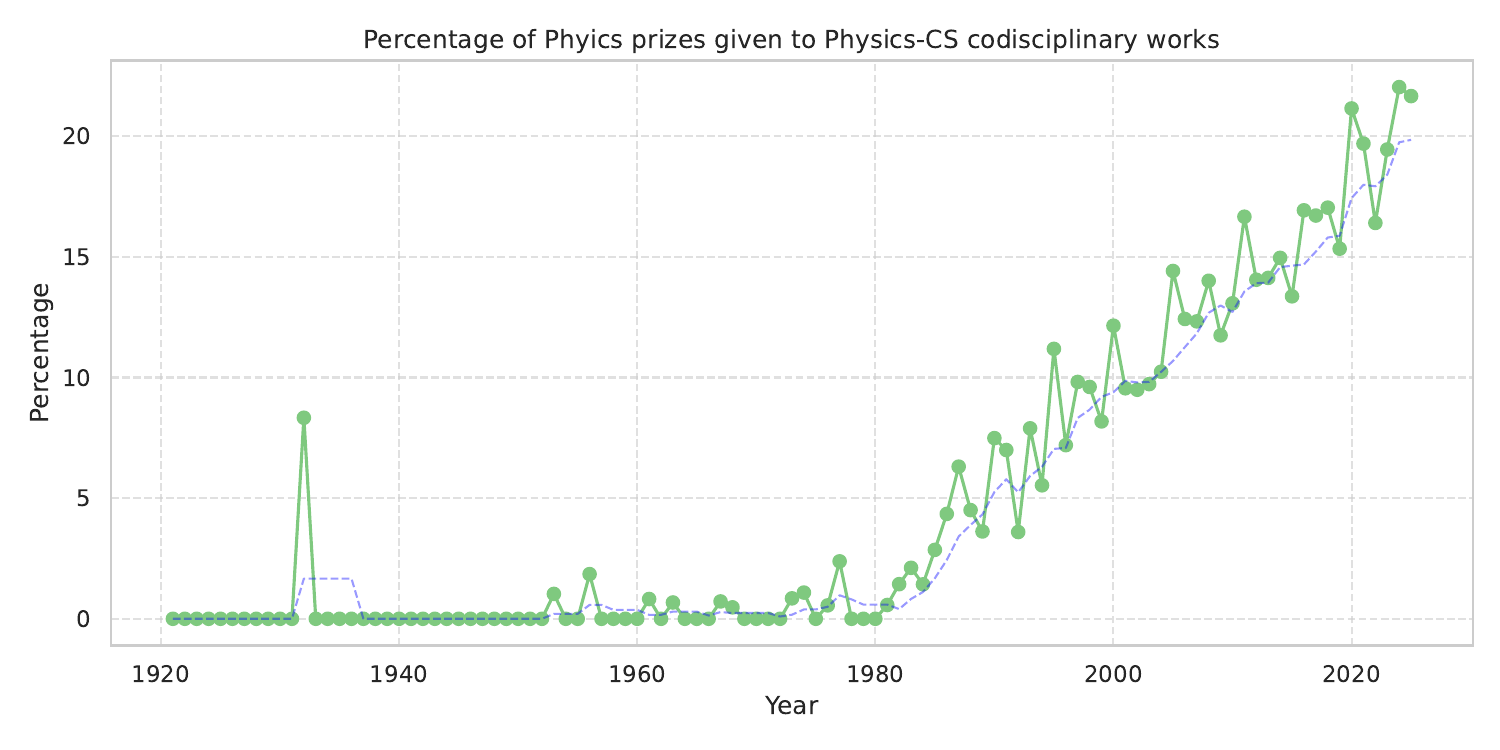}
        \caption{Timeline with CS}
        \label{fig:perc_phy_cs}
    \end{subfigure}
    \hfill
    \begin{subfigure}[b]{0.32\textwidth}
        \includegraphics[width=\textwidth]{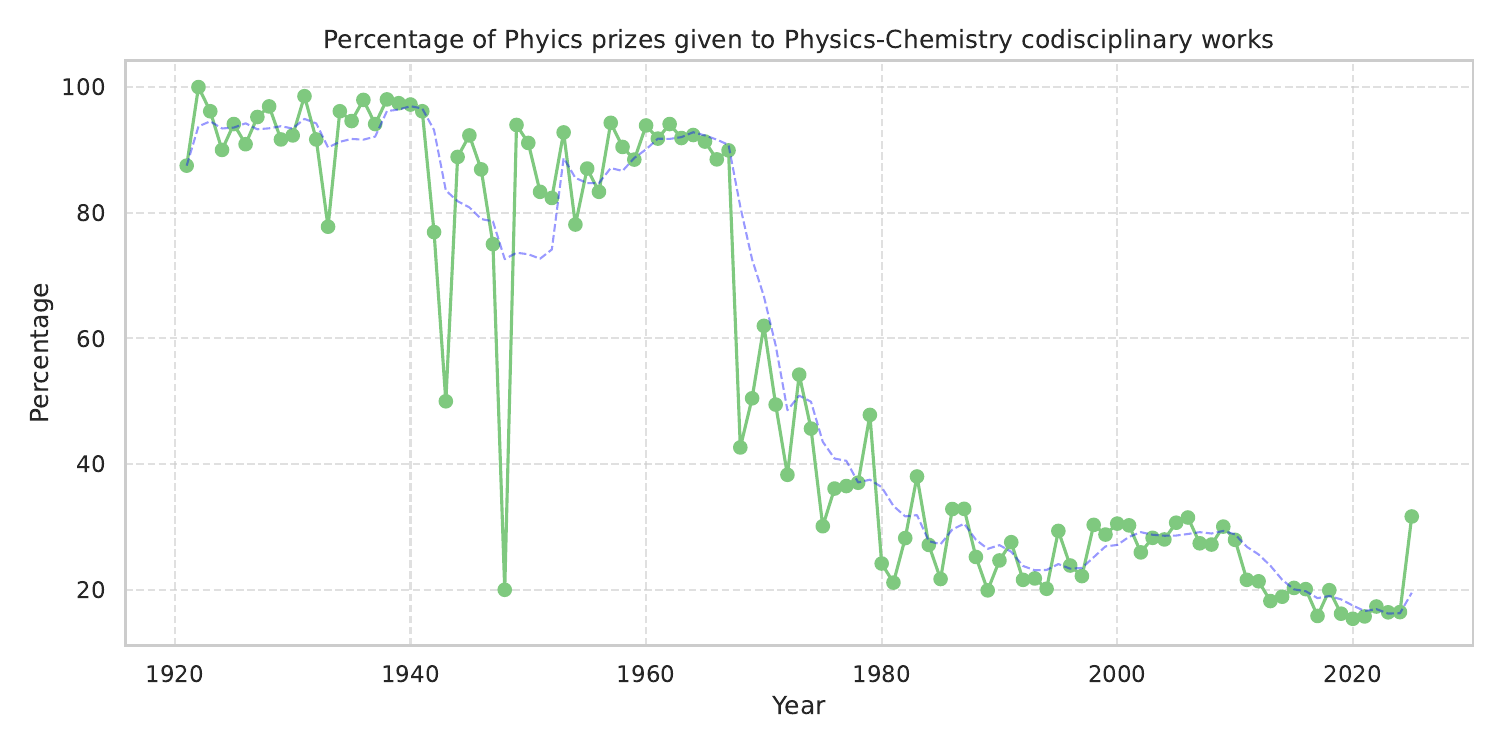}
        \caption{Timeline with Chemistry}
        \label{fig:perc_phy_chem}
    \end{subfigure}
    \hfill
    \begin{subfigure}[b]{0.32\textwidth}
        \includegraphics[width=\textwidth]{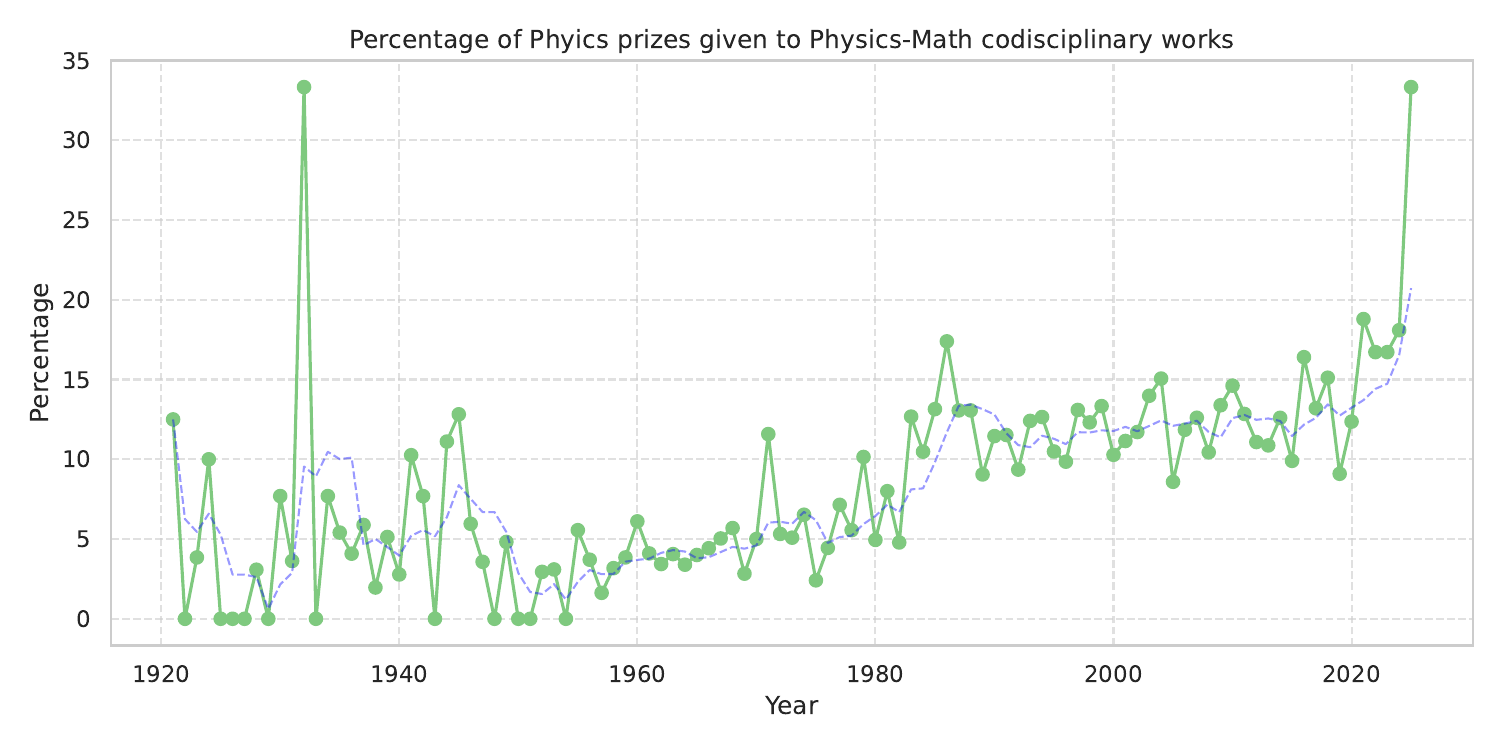}
        \caption{Timeline with Math}
        \label{fig:perc_phy_math}
    \end{subfigure}
    \caption{Comparison of Physics prizes. The blue dashed line in the percentage charts indicates the 5-year rolling average.}
    \label{fig:perc_comparison}
\end{figure}

We wanted to find out if the Nobel Prize in Physics going to a computer scientist was indicative of a wider trend of STEM research awards from different fields going to interdisciplinary or multidisciplinary computer science scholars. To answer this, we built a dataset of prizes awarded by computer science, physics, chemistry and mathematics societies in the USA and Europe, with CS and physics being the main focus of our analysis. We used OpenAI's GPT-4o with few-shot prompting to annotate each award instance as co-disciplinary or not. SI Section S3 describes the dataset we constructed and our approach to using GPT-4o.

For each prize in a given field, we wanted to find out if it had been awarded to a researcher for work in that field and a second field (say, an IEEE award being given for CS-Physics research), or not. For this purpose, we consider work in both fields to be interdisciplinary or multidisciplinary research or work that was a direct contribution to either field. We use the term 'co-disciplinary' as a shorthand term to describe such work. The results for Physics award shares to various co-disciplinary work in Computer Science, Chemistry and Math over time are shown in Figure \ref{fig:perc_comparison}.

\begin{figure}[t!]
\centering
\includegraphics[width=5in, height=3in]{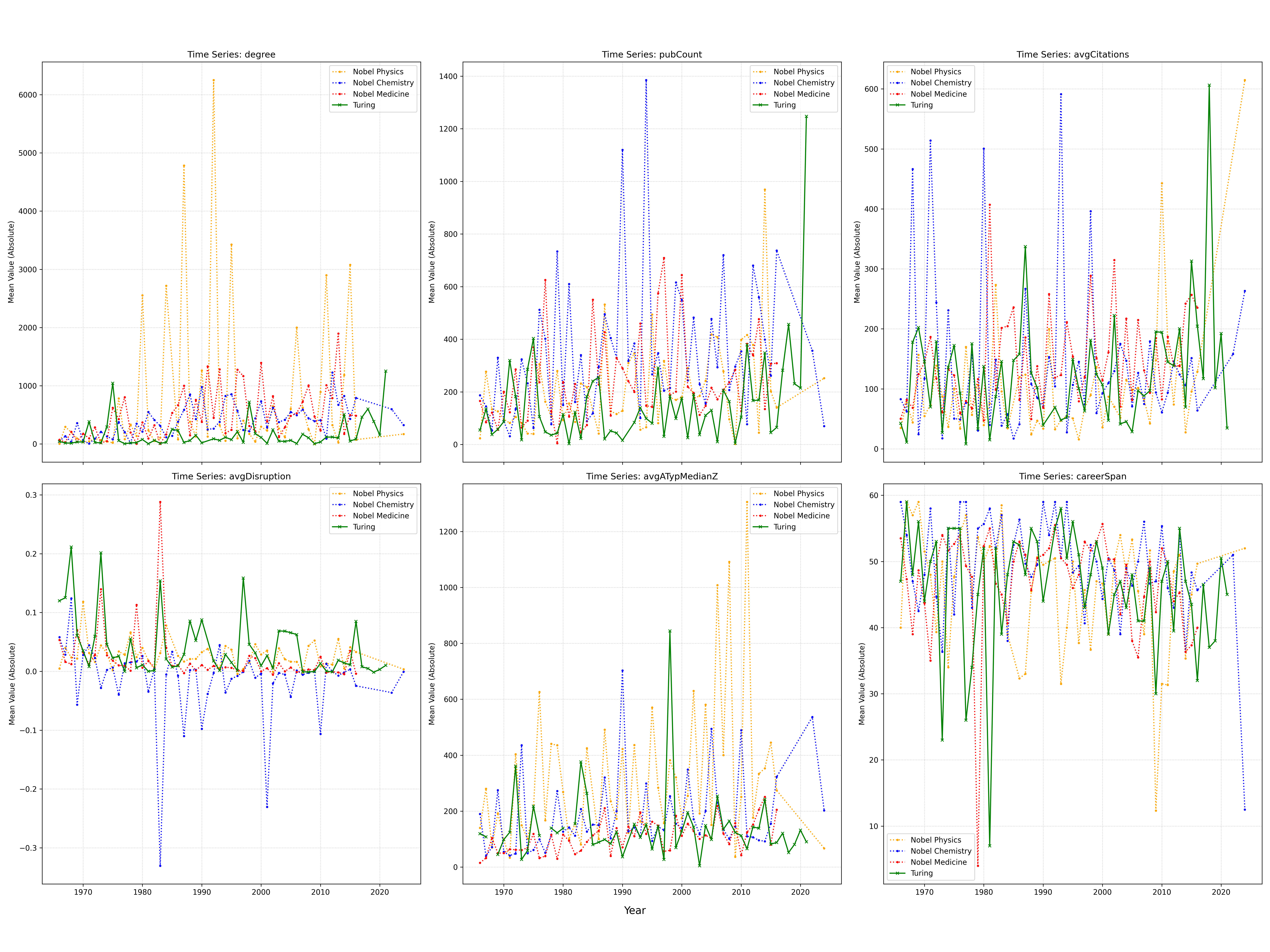}
\caption{\textbf{Time series graph of Nobel and Turing Award winner metrics since 1966.} Note that, if any award was shared in a given year, the pip for that award represents the average of that metric for the winning authors of that prize.}
\label{F:7}
\end{figure}

Our results show that a much larger proportion of Physics society awards go to co-disciplinary research compared to CS society awards. By the numbers, we find that 1,702 \((11.03\%)\) APS awards, 101 \((10.18\%)\) IOP awards, 87 \((8.29\%)\) Optica awards, 3 \((1.11\%)\) ASA awards, and 1 \((0.62\%)\) AVS award went to Physics-CS co-disciplinary work. Comparatively, only 263 \((3.71\%)\) ACM awards and 5 \((0.57\%)\) IEEE awards went to CS-Physics work. Major CS societies give out awards to in-discipline CS research at higher proportions than other fields, and that proportion has remained largely static over time. Meanwhile, the share of Physics, Chemistry and Math awards given to CS-adjacent research has increased steadily with over 20\% of physics society awards in 2024 going to physics-CS research.

\subsection{Comparison of Nobel Prize and Turing Award winners}

We performed a comparative study of Nobel Prize and Turing Award winners in SciSciNet using our graph database of author collaborations. We restricted this pool to 339 Nobel Prizes and 72 Turing Awards since 1966, the year the Turing began to be awarded. We found that Nobel Laureates had significantly larger co-author networks and larger publication counts, scoring above the mean for both, while Turing Award winners scored below the mean for those metrics. On the other hand, Turing winners appeared to have more impactful works based on less conventional sources, as demonstrated by their high average disruption score and lower than average Atypicality Median Z-score respectively, with Nobel Prizewinners displaying the opposite trend. These observations were found to be statistically significant. Recent work has explored a trade-off between academic novelty and 'superstar' collaboration which may explain the disruption score dynamic seen here \cite{A35}.

Fig. \ref{F:7} shows the trends for each metric in our comparative analysis for the three scientific Nobel Prizes and the Turing Award. Average citation count and average career span were similar across both sets. Our outcomes suggest that the work of Nobel Prizewinners is more prolific, with large co-author networks and publication numbers. This suggests that their work likely defines their field of specialization at large. Additionally, our results show that Turing Award winners may disrupt fields and build new areas of study with their work, reaching across multidisciplinary aisles more often than Nobel Prizewinners. 

\section{Discussion}
In science and academia, prizes considerably influence the direction of research and academic careers. This study investigates the relationship between prizewinning and educational collaboration in computer science. We examine how collaboration patterns change before and after prizewinning by analyzing data from 105 recognized scientific prizes awarded to 5046 scholars from 1965 to 2022. Our findings reveal that prizewinners in computer science are more likely to collaborate more frequently and earlier with other prizewinners than researchers who have not yet received similar recognition. In addition, they have a tendency to collaborate more with other prizewinners after receiving an award, a trend consistent across different age groups. Additionally, we observe that recipients of general prizes are more likely to collaborate than those of specialized prizes, but there are more specialized prizes than general prizes.

The field's pivotal role in contemporary scientific and technological advancement drives our choice to focus on computer science. With the rapid development of AI technologies, computer science has become a cornerstone of innovation, driving progress across numerous industries. The proliferation of research in this field has led to a substantial increase in publications and diverse collaboration patterns, providing a rich dataset for our analysis. Besides, computer science is characterized by its collaborative nature, with researchers frequently working together across institutions, countries, and disciplines. The complexity and interdisciplinary nature of modern computing problems necessitate collaboration, making it an ideal case study for understanding the impact of programming on academic partnerships. Furthermore, the field's dynamic and fast-paced environment offers a unique opportunity to observe how recognition and awards influence collaborative behaviors in real time.
 
Our study holds significant implications for the academic and research community. By uncovering the hidden relationship between prizewinning and collaboration, we provide valuable insights into how awards influence research trajectories and collaborative networks. These findings help scholars and institutions develop strategies to enhance their research impact and visibility. For individual researchers, understanding the tendency to collaborate more with other award winners post-prize can guide their collaboration choices and career planning. For academic institutions and funding bodies, these insights can inform the design of programs and incentives to foster productive and high-impact research collaborations.

Moreover, our findings have broader implications for the management and support of scientific research. We advocate for policies that encourage and facilitate partnerships between emerging and established researchers by highlighting the importance of prizewinning in shaping collaborative patterns as a path to a more interconnected and dynamic research environment, accelerating scientific progress and innovation.

\section{Methods}
\subsection{Coarsened Exact Matching}
CEM \cite{A28} is a method which prunes data points to achieve a more similar distribution of covariates between the treatment and control groups, also known as balancing. It is different from other matching methods as it coarsens continuous variables such that close values are assigned the same numerical value before performing exact matching. The method allows each variable to be coarsened to different extents, and we use a greedy algorithm to determine the best extent of coarsening for each covariate to achieve a good matching outcome overall.

We built a graph database of prizewinning authors and their co-authors using Neo4j. Across 105 awards, we found $5,947$ unique prizewinning authors in the SciSciNet dataset with a total of $968,936$ non-prizewinning co-authors between them (This data set contains on average $~163$ co-authors per prizewinner). We selected five attributes for authors that would be representative of each author's career. The attributes are \textbf{degree} (number of co-authors), \textbf{Pub Count} (number of papers associated with the author in SciSciNet), \textbf{Avg. Citations} (mean of citations across those papers), \textbf{Career Span} (number of years between the author's first and last paper) and \textbf{Avg. Disruption} (mean of disruption across all papers). Disruption score is a value provided for many papers in the SciSciNet dataset which "quantifies the extent to which a paper disrupts or develops the existing literature" calculated based on the paper's citation networks. Importantly, the pre-award collaboration objective requires that all these attributes are calculated only for the period before the prizewinners win their first prize, and in the case of non-winners, up to the fifth year of their collaboration with a prizewinner. The rationale for computing the non-winner attributes this way is described in SI Section S3. 

\subsection{Regression Models}
For the weighted regression model, we first standardize the control variables to have a mean of 0 and a standard deviation of 1. This ensures that the covariates with larger scales do not overshadow the smaller ones. We then fit the model, including all five covariates. The dependent variable is whether an author won a prize (1/0). For whole-career collaboration, the treatment is a strong collaboration over one's career with a different prizewinner with one's own odds of winning. Note that for the pre-award objective, the treatment is based on pre-award metrics for prizewinners or equivalent 5-year collaboration metrics for non-prizewinners. Control variables were standardized before inclusion in the model.

\section{Acknowledgments}
The authors thank Dr. Ching Jin from the University of Warwick for helpful discussions and advice during the early preparation of this manuscript.

\bibliography{main.bib}\

\end{document}